\documentclass[%
 reprint,
superscriptaddress,
 amsmath,amssymb,
 aps,
prb,
showkeys
]{revtex4-1}
\pdfoutput=1
\usepackage{placeins}
\usepackage{graphicx}
\usepackage{comment}
\usepackage{amsmath}
\usepackage{textcomp}
\usepackage{gensymb}
\usepackage[caption=false]{subfig}
\usepackage{framed}
\usepackage[normalem]{ulem}
\usepackage[export]{adjustbox}
\usepackage{dcolumn}
\usepackage{bm}
\usepackage{array,ragged2e}
\usepackage{lettrine}
\usepackage[dvipsnames]{xcolor}
\usepackage{multirow}

\newcolumntype{P}[1]{>{\Centering\hspace{0pt}}p{#1}}
\begin{document}
\preprint{APS/123-QED}
\title{XeF$_2$-Enhanced Focused Ion Beam Etching and Passivation of GaSb}

\author{Pierce~Maguire}
\affiliation{School of Physics, Trinity College Dublin, Dublin 2, Ireland, D02 PN40}
\affiliation{AMBER Centre, CRANN Institute, Trinity College Dublin, Dublin 2, Ireland, D02 PN40}

\author{Darragh~Keane}
\affiliation{AMBER Centre, CRANN Institute, Trinity College Dublin, Dublin 2, Ireland, D02 PN40}
\affiliation{School of Chemistry, Trinity College Dublin, Dublin 2, Ireland, D02 PN40}

\author{Brian Kelly}%
\affiliation{Eblana Photonics Ltd., 3 West Pier Business Campus, D\'u{}n Laoghaire, Dublin, Ireland, A96 A621}

\author{Colm C. Faulkner}
\affiliation{AMBER Centre, CRANN Institute, Trinity College Dublin, Dublin 2, Ireland, D02 PN40} 

\author{Hongzhou~Zhang}
\email{Corresponding author E-mail: Hongzhou.Zhang@tcd.ie}
\affiliation{School of Physics, Trinity College Dublin, Dublin 2, Ireland, D02 PN40}
\affiliation{AMBER Centre, CRANN Institute, Trinity College Dublin, Dublin 2, Ireland, D02 PN40}

\newcommand{\note}[1]{\color{Red}\textbf{** #1 **}\color{black}}
\newcommand{\reply}[1]{\color{Green}\textbf{** #1 **}\color{black}}
\date{\today}
\begin{abstract}
Focused ion beams (FIBs) are widely used to modify optoelectronic devices, but their utility is severely restricted in some III-V materials due to adverse effects such as nanofibre growth and contamination. This study describes an effective machining method for the modification and passivation of GaSb using a focused Ga$^+$ beam with the addition of XeF$_2$ gas. The added gas suppressed the nanofibre growth and allowed for controlled etching. Moreover, the milled surface exhibited much lower roughness and contained much lower levels of carbon and oxygen contaminants. These effects are attributed to the formation of a gallium fluoride layer, which inhibits the catalytic vapour-liquid-solid (VLS) nanofibre growth. The fluoride layer prevents the surface oxidation, acting as an effective passivation layer. This gas-assisted FIB process eliminates the idiosyncratic challenges of modifying GaSb with an ion beam and it can be implemented cost-effectively and rapidly in conventional FIB systems---allowing for site-specific modification of bespoke GaSb-based devices. 
\end{abstract}

\keywords{GaSb, XeF$_2$, VLS growth, Focused Ion Beam, Passivation Layer}
\maketitle

\section{Introduction}
The III-V semiconductor compound GaSb has a relatively small bandgap of 0.7 eV and is the basis of many photovoltaic and thermophotovoltaic (TPV) systems \cite{Chen1991a}, particularly suitable for infrared, optical and high speed devices in the spectral region from 1.3 to 6.5 $\mu$m \cite{Bagheri2013,Wang1996}. Controlling the morphology of III-V laser diode devices can be critical to manipulating their characteristics, including threshold currents and wavelength
\cite{Mack1998,Ross2006,Vallini2009}, for example, single-mode lasers based on machined surface gratings. Minimising roughness---for instance at mirror surfaces---is of importance for preventing scattering and the associated reduction in optical power.

Conventional machining methods for GaSb typically include combinations of lithography with plasma or reactive ion etch (RIE) exposure \cite{Giehl2003}. Halogen and hydrogen plasmas are effective at producing smooth features with an etching rate of $\sim 10^2 -10^3$ nm/min and a roughness of several nanometers \cite{chang1982}. However, they are limited by the need for masking, since masks are typically manufactured only for large, wafer-level scaling. For small scale production---particularly of prototype or bespoke devices---manufacturing using lithographic mask plates and transfer is prohibitively slow and expensive. Although polymer photo-resists are one potential alternative, they are unsuitable for use in processed devices---for example in failure analysis or post-production modification---as these can introduce contaminants and require the application of solvents in their removal---risking damage to functioning devices.

Focused ion beams (FIBs) are particularly useful in the fabrication and modification of semiconductor devices given the throughput, resolution and customisability that they offer. However, GaSb poses idiosyncratic challenges for machining using conventional Ga$^+$ FIB. Many reports have demonstrated either a porous, honeycomb-like structure or nanofibre growth evolving under the beam \cite{Schoendorfer2006, Lugstein2007a,Kluth2005,Perez-Bergquist2009,Datta2014}. \citeauthor{Schoendorfer2006} proposed a vapour-liquid-solid (VLS) growth mechanism for nanofibres under the Ga$^+$ ion beam, involving the formation of Ga droplets on the surface of GaSb \cite{Schoendorfer2006}. Other ions such as Au$^+$ and Sn$^+$ also cause the growth of undesired nanofibres \cite{Nitta2002,Lugstein2007a,Schoendorfer2006,Perez-Bergquist2009,Datta2014,El-Atwani2016}. Many other ions (especially light species) cause swelling of the material and subsurface cavity formation \cite{Stanford2017a, El-Atwani2016}.

Another challenge in GaSb modification is surface oxidation---being much more likely to occur than in GaAs or InP---as the reaction is not self-limiting. The surface created by an ion beam seems to be particularly vulnerable to oxidation \cite{Datta2014}, meaning that passivation strategies must be employed on the surface to ensure sufficient longevity of devices \cite{Mizokawa1988, PapisPolakowska2006a, Spicer1979, McDonnell2011}. Passivation of the pristine GaSb surface has been performed using several wet chemical methods, including with sulfur and ruthenium \cite{Lin1998,Dutta1994,Dutta1995,Dutta1994}. However, wet chemical methods are unsuitable for the passivation of FIB-modified GaSb surfaces in functioning devices. Since the performance of GaSb devices is closely linked to the quality of the machined surface \cite{Dutta1997,PapisPolakowska2006a}, these adverse secondary effects render current FIB machining processes impractical for use with GaSb. 

In this paper we show that these effects of FIB applied to GaSb can be effectively mitigated by the application of XeF$_2$ gas to the sample surface during ion irradiation. This gas-assisted etching approach has been used in FIBs before on SiO$_2$ \cite{Utke2008, Ebm2010} and GaN/GaN-Si$_3$N$_4$ \cite{Ross2006, Mitrofanov2017} to enhance the etching rate and reduce redeposition. A similar approach with an electron beam has been successful in GaAs, albeit with a much lower throughput \cite{Ganczarczyk2011}. We show that in GaSb, it allows the combination of the spatial resolution and versatility of focused ion beam machining with the smooth surface finish of plasma etching. Milled features are deeper, smoother and contain vastly less contamination with C and O species than with the Ga$^+$ beam alone. This method can be used for much more effective prototyping of novel GaSb-based devices. 

\section{Experimental Details}
A single crystal wafer of n-doped GaSb (110) was made by \textit{Wafer Technology Ltd}. and provided by \textit{Eblana Photonics Ltd}. Focused ion beam etching was performed with 30 keV Ga$^+$ at a normal angle of incidence in a \textit{Zeiss Auriga} dual beam FIB system. The gas injection system (GIS) nozzle was placed $\sim$100 $\mu$m above the sample surface and used to inject XeF$_2$ at 5\degree{}C in the direction of the sample surface. The vacuum level at normal equilibrium was $\sim$2$\times$10$^{-6}$ mbar which rose to $\sim$1$\times$10$^{-5}$ mbar with the gas valve open.  

Squares of 10 $\mu$m $\times$ 10 $\mu$m were irradiated with a beam current of 20 pA. The beam was scanned rapidly (20000 Hz in x and 20 Hz in y) across the selected regions. 
Scanning electron microscopy (SEM) was performed both in-situ in the FIB system and ex-situ in a \textit{Zeiss Ultra Plus} and a \textit{Zeiss Supra} SEM with a \textit{Bruker Nano XFlash 5030} detector with which EDX hypermaps were obtained using a range of electron beam energies at an angle of 0\degree{}.
Atomic force microscopy was performed in an \textit{Oxford Asylum MFP-3D} system in atmospheric conditions in tapping mode. The probes used had a resonance frequency of 300 kHz and a force constant of 40 N/m (\textit{Budget Sensors tap300Al-G}).
Raman Spectroscopy was performed in a \textit{Horiba Labram Aramis} Raman spectrometer with a 785 nm laser. A 1800 lines/mm diffraction grating and a 100$\times$ objective aperture (NA=0.9) were utilised (laser spot size was $\sim$1 $\mu$m). These spectra were averages, comprised of 20 acquisitions, each of 1 s duration at a single point. A power of less than 1 mW was ensured to minimise damage to the sample. 
\begin{figure*}[ht!]
\centering
\captionsetup[subfigure]{labelformat=empty}
\subfloat[]{\label{sub:semafm}\includegraphics[height=2.4in]{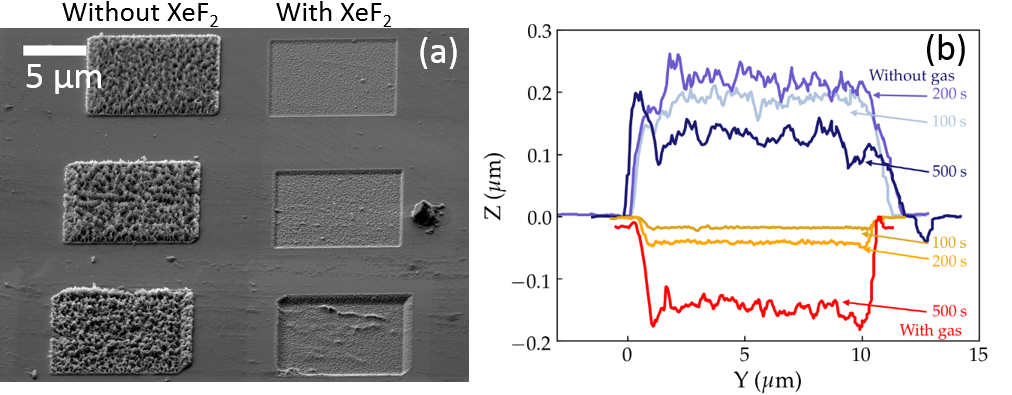}}\quad
\caption{SEM and AFM characterisation of 30 keV Ga$^+$ irradiated squares. (a) shows the six 10 $\mu$m $\times$ 10 $\mu$m squares, imaged with a beam energy of 5 keV and at an angle of 54\degree{} to highlight the feature depth and morphology. The left and right columns were milled without and with the XeF$_2$ gas respectively. From the top, the rows were irradiated for 100, 200 and 500 s respectively. (b) shows AFM line profiles of the same irradiated squares averaged over the width of each for the `with gas' and `without gas' cases for the 100 s, 200 s and 500 s irradiations.
}
\label{fig:AFMsem}
\end{figure*}
\section{Results}
Figure \ref{fig:AFMsem}(a) shows an SEM image of the GaSb surface with 10 $\mu$m $\times$ 10 $\mu$m irradiated regions. There are six such squares in two columns, imaged at an angle of 54\degree{}. The left and right columns were irradiated with the XeF$_2$ gas valve closed and open respectively. The first, second and third rows were irradiated for 100, 200 and 500 s respectively i.e. doses of 1.25, 2.5 and $6.25\times 10^{16} \mathrm{\ Ga}^+ \mathrm{\ cm}^{-2}$. There is a clear difference in morphology apparent between the two columns. Shadowing effects in the image suggest that in the `without-gas' squares the height of the irradiated surface is increased in comparison to the surrounding areas, while the height of the `with gas' squares appears to be reduced, particularly as the ion dose increases. 
\begin{figure*}
\centering
\captionsetup[subfigure]{labelformat=empty}
\subfloat[]{\label{sub:allsem}\includegraphics[height=3.5in]{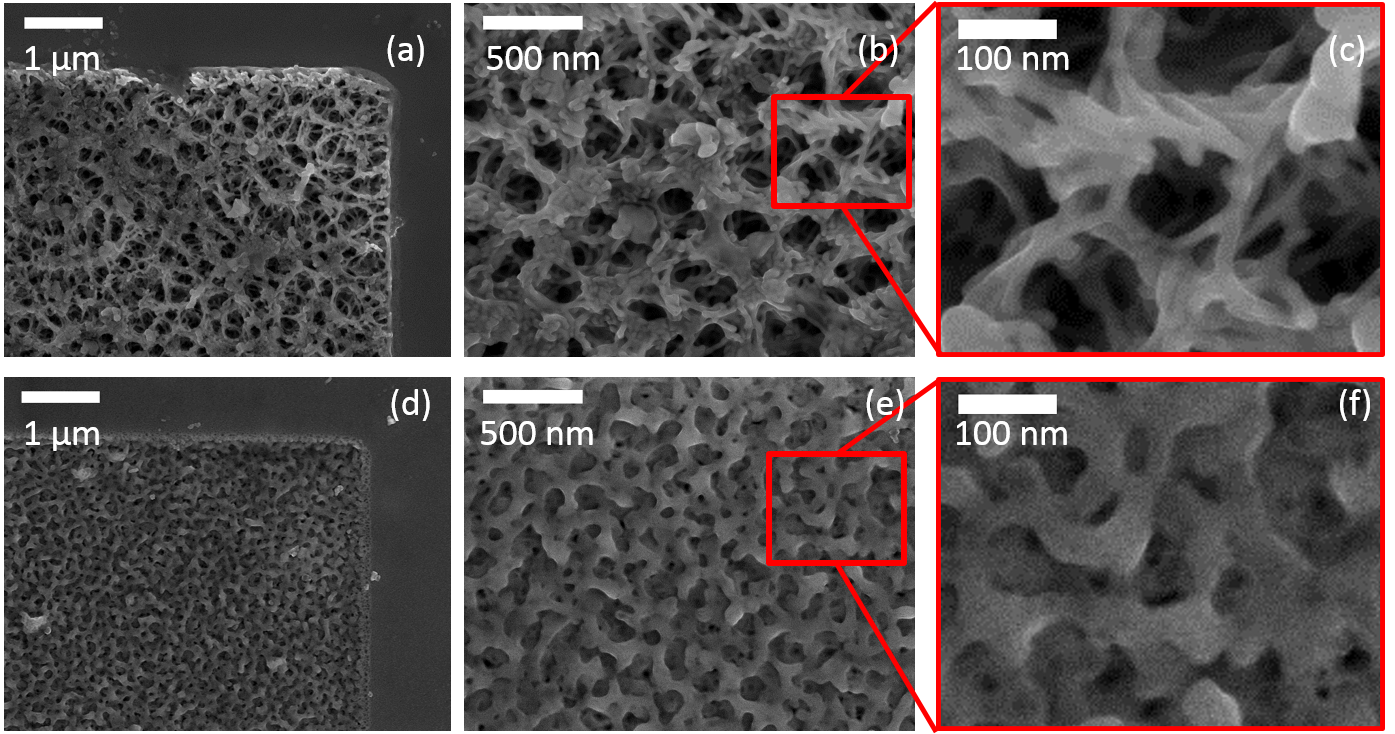}}\quad
\caption{SEM images acquired at a stage angle of 0\degree{} of 100 s irradiated squares. (a),(b) and (c) show squares irradiated without gas with increasing magnification respectively.  (d),(e) and (f) show squares irradiated with gas with increasing magnification respectively.
}
\label{fig:sem}
\end{figure*}
\begin{table}
    \centering
    \begin{tabular}{ c | c | c |c |c}
    \hline\hline
     \multirow{2}{*}{Time (s)} & \multicolumn{2}{c|}{RMS (nm)}& \multicolumn{2}{c}{$\Delta$H (nm)}\\ \cline{2-5}
        & With XeF$_2$ & No XeF$_2$ & With XeF$_2$  & No XeF$_2$ \\    \hline
      100 & 6.7 & 71.8 &-18 &188\\
      200 & 14.5 & 99.6 & -42 & 222\\
      500 & 45.9 & 109.3 & -142 & 127 \\ \hline
     
    \end{tabular}
    \caption{Summary of key parameters from the AFM measurements including irradiation time, use of XeF$_2$ gas, RMS roughness, and average change in height ($\Delta$H)}.
    \label{table:afm}
\end{table}

Figure \ref{fig:sem} shows a more thorough SEM study of the modified GaSb surface, irradiated for 100 s both with (a)(b)(c) and without (d)(e)(f) gas. Images of the 200 s and 500 s regions are shown in the supplementary information. The ion irradiation causes the growth of a rough, porous network of fibres. The width of the nanofibres is reasonably uniform, varying between $\sim20$ and $\sim30$ nm. They are observed to overlap substantially so that relatively large pores (some $>100$ nm) are apparent. Particle-like features are also observed among the nanofibres which may be GaSb crystallites \cite{Kluth2005}. With the addition of the XeF$_2$ gas, nanofibre growth is not apparent and the pores are now much smaller ($\sim30$ nm) while also appearing to be much more shallow. The irradiated surface appears much less rough, in good agreement with the AFM. 

\begin{figure}
\centering
\subfloat[Non-Irradiated]{\label{sub:EDXSpectra}\includegraphics[height=5.15in]{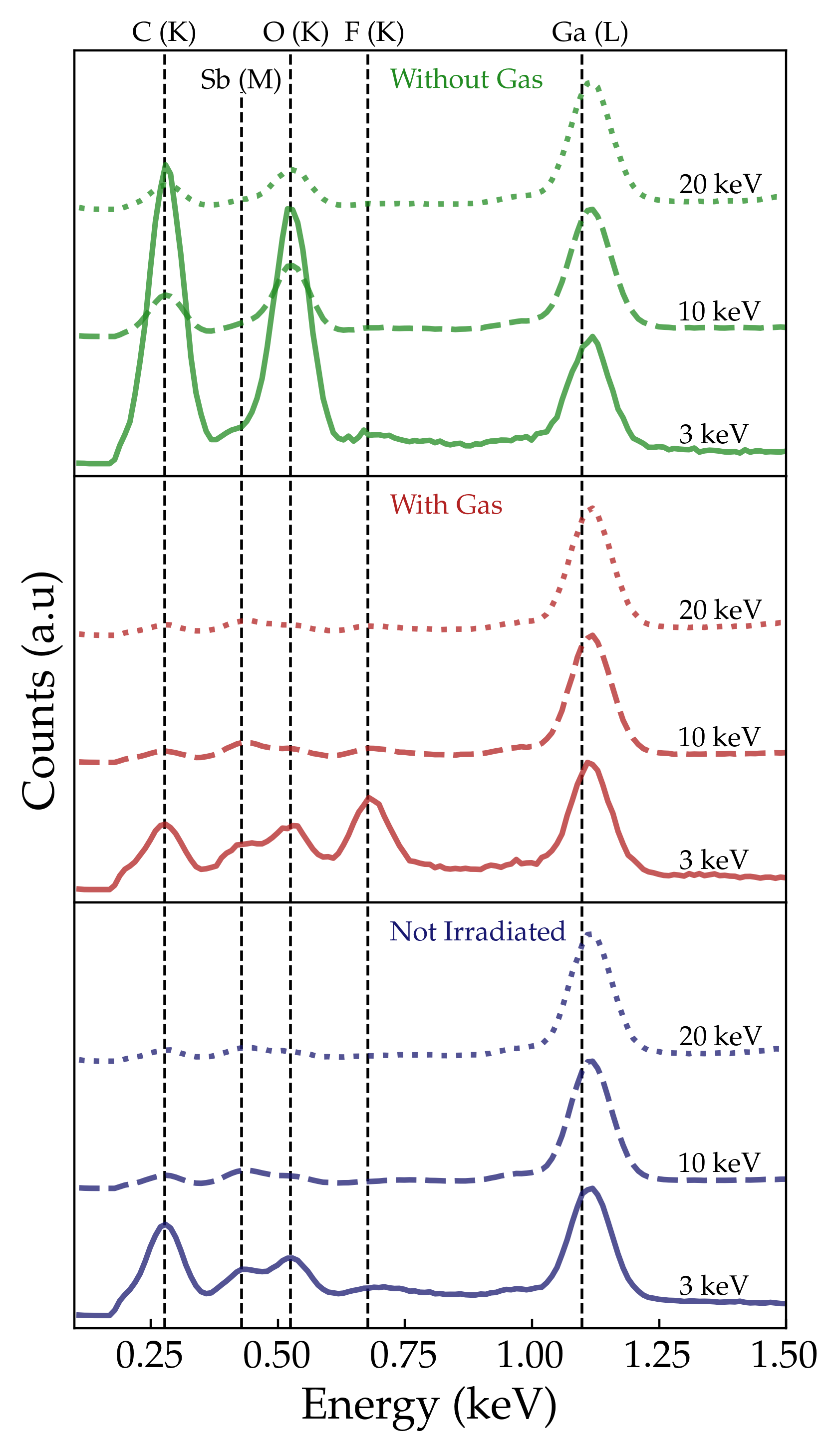}}
\caption{EDX spectra of the non-irradiated (blue), irradiated `with gas' for 100 s (red) and irradiated `without gas' for 100 s cases respectively for 3 keV, 10 keV and 20 keV electron excitations (solid, dashed and dotted). The intensity of each spectrum is normalised to the maximum of the Ga-$L$ peak.
}
\label{fig:edx}
\end{figure}

\begin{figure*}
\centering
\captionsetup[subfigure]{labelformat=empty}
\subfloat[][]{\label{sub:edxmaps}\includegraphics[height=2.00in]{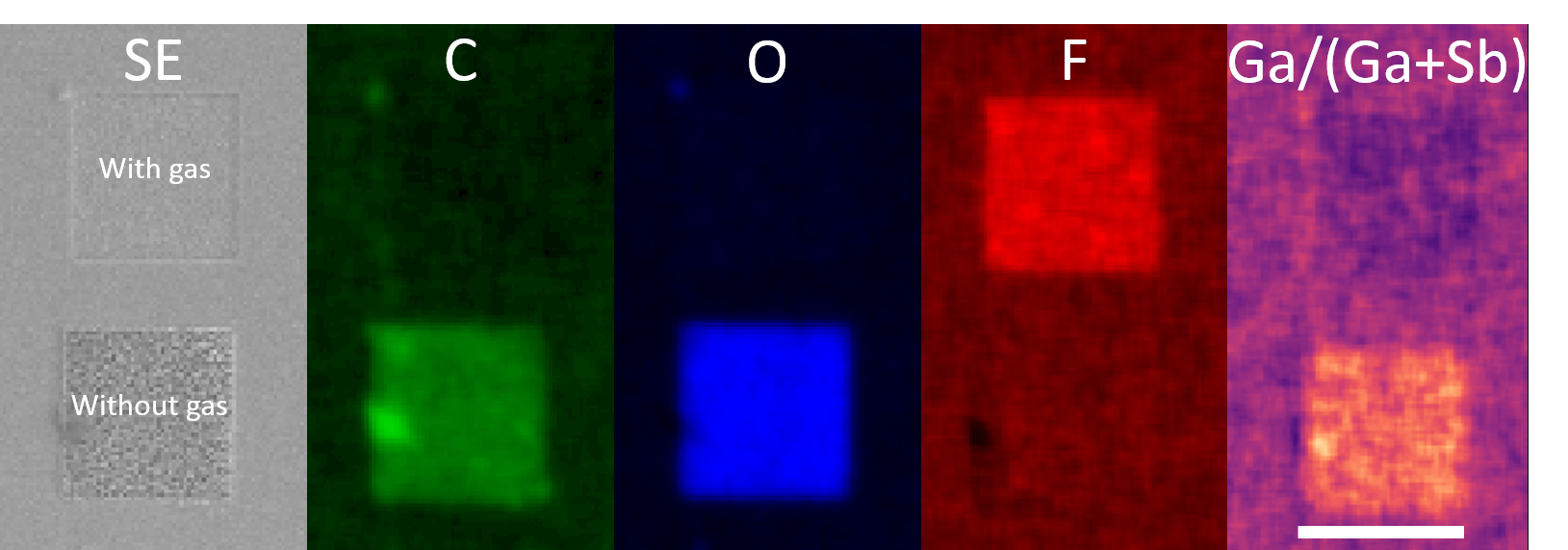}}\qquad
\caption{Intensity maps of the 100 s irradiated regions acquired at 10 keV. As labelled, they are a secondary electron image (in grayscale) and EDX maps of the carbon (green), oxygen (blue) and fluorine (red) K lines as well as a fraction map of Ga-$L$/(Ga-$L$+Sb-$L_\beta$). The Ga-$L$ and Sb-$L_\beta$ maps are provided in the supplementary information. The scale bar is 10$\mu$m.}
\label{fig:edxmaps}
\end{figure*}

Figure \ref{fig:edx} shows EDX spectra of the non-irradiated (blue), 100 s irradiated with gas (red) and 100 s irradiated without gas (green) regions. The spectra were acquired using beams of 3, 10 and 20 keV electrons (solid, dashed and dotted lines respectively) for excitation. With data acquired using different beam energies, different signal depths are probed which allows comparison between the near-surface region and the bulk. For the non-irradiated GaSb, the 20 keV and 10 keV spectra appear as expected---with the clear presence of Ga and Sb and small amounts of O and C. As the beam energy decreases to 3 keV, the intensity of the O-$K$ and C-$K$ peaks increases sharply, indicating that these contaminants are strongly present near the surface. After irradiation for 100 s with the Ga$^+$ beam alone, there is a large increase in the O and C signals in all three spectra, again this is especially clear in the 3 keV spectrum. This suggests that the surface irradiated without gas---which we know is now highly porous and composed of nanofibres---becomes highly oxidised while also becoming coated with hydrocarbon material. By contrast, the spectra of the region irradiated in the presence of XeF$_2$ gas show very little O and C. All three spectra acquired of the `with gas' case are very similar to their starting material counterparts---demonstrating very low O and C content. However, there is one notable difference which is the new presence of a large F-$K$ peak, clearest in the 3 keV case. As with the O and C signals previously, this peak diminishes rapidly for the higher energy electron beams, suggesting that F is embedded at or near the surface. 

These results are highlighted by EDX maps of the 100 s irradiated regions which were excited by the 10 keV electron beam, shown in Fig. \ref{fig:edxmaps}. The first map is a secondary electron image showing both 100 s regions (which are labelled) and the surrounding substrate. The carbon and oxygen maps are shown in green and blue respectively and illustrate clearly a much higher signal in the rough, `without gas' region than in either the surrounding area or the `with gas' region. The fluorine map is shown in red and shows a much higher intensity for the irradiated `with gas' region than either the `without gas' region or the surrounding area. The final map is of Ga/(Ga+Sb), showing a significantly higher Ga composition in the `without gas' region caused by the implantation of excess Ga from the beam. The `without gas' region is much more similar to the surrounding material. The EDX characterisation has shown that under irradiation in the presence of XeF$_2$, contamination is suppressed while F atoms are embedded in the surface.   

\begin{figure*}
\centering
\subfloat[Without gas]{\label{sub:GaSbSpectraWithout}\includegraphics[height=2.5in]{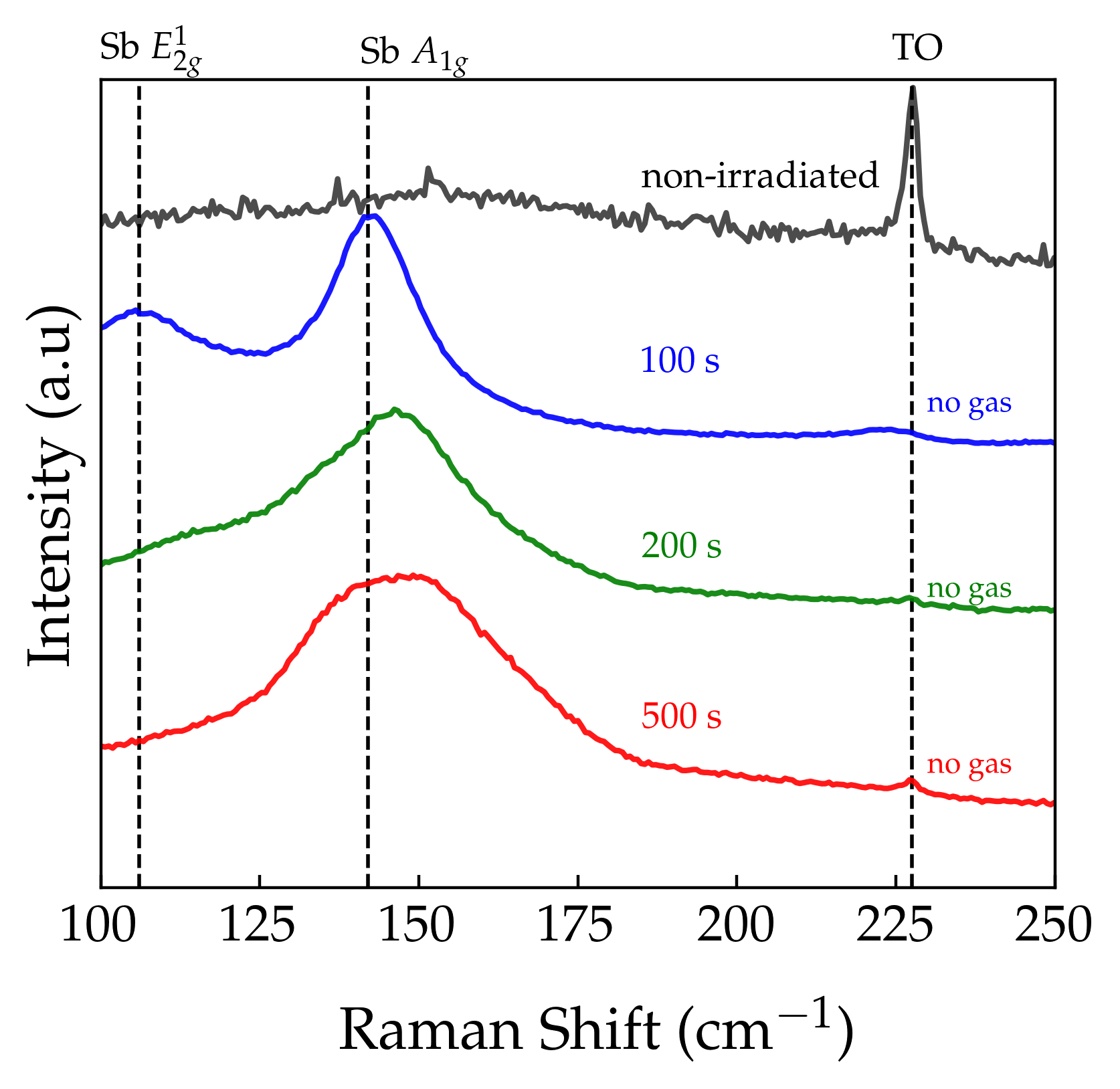}}\quad
\subfloat[With gas]{\label{sub:GaSbSpectraWith}\includegraphics[height=2.5in]{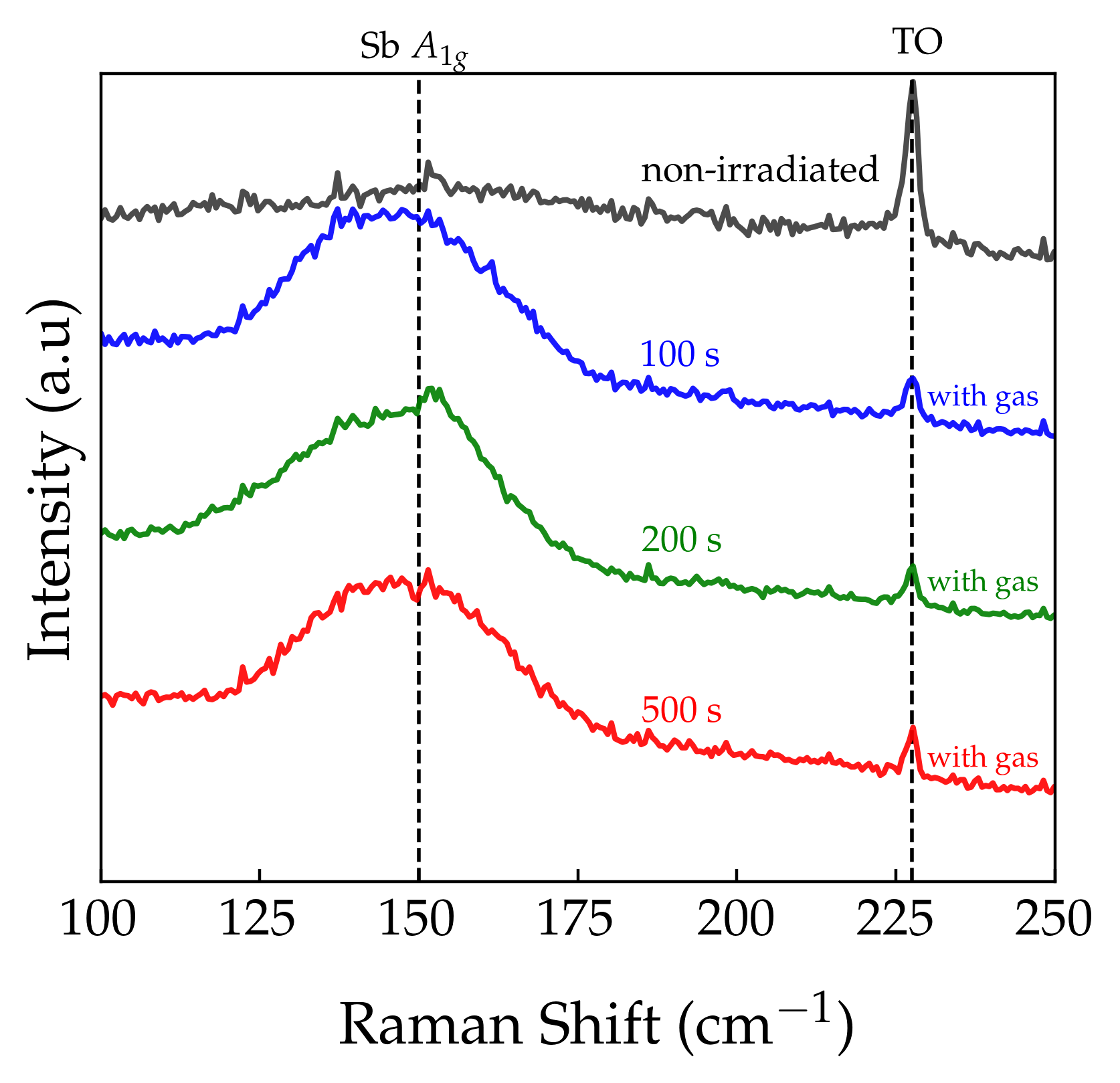}}\quad
\caption{
(a) and (b) show Raman spectra of GaSb before and after irradiation (`without gas' and `with gas' respectively) with Ga$^+$ at 30 keV with a 0$^{\circ}$ angle of incidence. The evolution of the spectra with increased irradiation time is shown descending from the top. The samples were excited with a 785 nm laser. The TO GaSb and the Sb $A_{1g}$ peaks are labelled accordingly and marked with the vertical dashed lines. The intensity is normalised to the maximum of each spectrum.}
\label{fig:GaSbSpectra}
\end{figure*}

The results of Raman spectroscopy of the same regions are presented in figure \ref{fig:GaSbSpectra}(a),(b) for the `without gas' and `with gas' regions respectively. The spectrum acquired for the non-irradiated GaSb is presented in both graphs for comparison. The TO peak in the non-irradiated GaSb is clearly visible at $\sim$227.5 cm$^{-1}$---in good agreement with literature for the (110) orientation \cite{DiasDaSilva1995, Kim1993}. 

In all three regions irradiated without gas (see Fig. \ref{fig:GaSbSpectra}(a)), the TO peak intensity is very low and close to undetectable. With a laser wavelength of 785 nm, Raman signal depth in GaSb is estimated to be $\sim$90 nm \cite{Munoz1999, Maslar2007}. Since the AFM, SEM and EDX results indicate an increased height ($\sim 200$nm) composed of redeposited material and contaminants, the Raman signal volume should be contained within the material near the surface. We therefore conclude that the elevated, nanofibrous surface created by the ion beam alone contains trace amounts of crystalline GaSb. For the `with gas' case (see Fig. \ref{fig:GaSbSpectra}(b)), all three spectra show a more moderate drop in TO intensity. Since we know from the SEM and AFM data that the buildup of nanofibrous material at the surface is suppressed in these cases, the loss of signal is attributed to amorphisation of the GaSb crystal. TRIM calculations (see Fig. S3) indicate that damage due to irradiation with 30 keV Ga$^+$ is occurring up to a depth of $\sim$50-60 nm \cite{ziegler2010srim}, occupying most of the Raman signal volume. Additionally, the TO peak is not observed to down-shift, as expected in defective, crystalline GaSb. Therefore, we can also conclude that the boundary between amorphous and crystalline material is relatively abrupt in all cases, with very little defective-but-crystalline material present.

There are other noteworthy features in the Raman spectra of irradiated regions, including a broad feature centred in the region 144-153 cm$^{-1}$. We attribute this to the A$_{1g}$ peak of elemental Sb which has a larger cross-section than GaSb \cite{Zhou2011,Lafuente2015, Kim1993}. Although the Sb A$_{1g}$ peak may be very weakly present in the non-irradiated spectra, it becomes much more pronounced in all irradiated cases. In the spectra from regions irradiated with gas, the results are very similar to each other and this broad feature dominates. The E$^1_{2g}$ peak is also discernible, strongest in the 100 s `without gas' case \cite{Campos2006}. With increasing irradiation time in the `without gas' cases, both Sb peaks are observed to upshift and broaden. For the 500 s case, the peaks cannot be resolved from one broad feature. This seems consistent with the spectra becoming dominated by Sb of worse crystallinity as irradiation time increases \cite{Roy1996}.
From Raman spectroscopy, clear effects of irradiating in the presence of XeF$_2$ gas are to produce a reduced height of amorphous material and reduced contamination between the surface and the crystalline surface beneath. 

\section{Discussion}
The melting point of pure, metallic gallium is $\sim$30\degree{}C. It is widely established---and reported in their respective phase diagrams---that liquid Ga droplets form at similar temperatures when Ga is in excess in non-stoichiometric gallium antimonide or gallium arsenide \cite{Hansen1958, Youdelis2007, Okamoto1992}. A conventional Ga$^+$ FIB is known to both implant excess gallium and locally increase the specimen temperature, causing the formation of these droplets. Our EDX results certainly show that the ion irradiation adds excess Ga atoms. The resulting droplet formation has been mitigated---but not eliminated---in another work by controlling the temperature in a cryo-FIB system \cite{Santeufemio2012}. However, in most experiments these droplets catalyse the VLS growth of the undesired, amorphous nanofibres during the non-equilibrium state caused by ion irradiation \cite{Lugstein2007a}.

From the AFM and SEM data it is clear that the gas prevents the growth of nanofibres under the beam. From the EDX data, the addition of XeF$_2$ gas has contributed two principle causes for this: the removal of excess Ga, and the implantation of F atoms. We suggest that both effects are necessary to prevent droplet formation and hence nanofibre growth. While the removal of excess Ga atoms plays an important role in maintaining the stoichiometry, there is another possible source of Ga. Coalescence of the V element in III-V materials has been reported previously \cite{Farrow1977, Campos2006, Alarcon-Llado2013}, particularly of Sb in GaSb after oxidation \cite{Farrow1977} or ion irradiation \cite{Perez-Bergquist2009, Zhou2011}. The coalescence of Sb can lead to Ga segregation at the surface \cite{Perez-Bergquist2009} and our Raman spectra indicate the presence of metallic Sb in both the `with gas' and `without gas' cases. The implantation of F atoms is therefore critical to passivating Ga atoms at the surface. 

Given the lack of nanofibre growth or O/C atom contamination at the surface, we conclude that the F-passivated surface is chemically and thermally stable. The reaction of XeF$_2$ with GaAs to form GaF$_3$ at the surface has been reported previously \cite{Varekamp1994, Nienhaus1996, Simpson1996}. An analogous reaction is likely to be occurring under the ion beam \cite{Youdelis2007, Okamoto1992}. 

The outcome of this process is that etched GaSb surfaces can be produced (when compared with conventional FIB) with much better throughput, much lower roughness, more spatial confinement  and more chemical and thermal stability---while maintaining all of the site specificity that is advantageous over lithography techniques. Many experimental parameters remain to be tested and optimised together such as beam energy, beam current, gas vapor pressure/flow rate, specimen temperature, scanning strategy, angle of incidence and more. Considerable improvements remain to be made here but already this process should be of considerable utility in the modification of manufactured devices for bespoke applications.

\section{Conclusion}
In summary, we report an effective nanofabrication method for GaSb-based systems. The addition of XeF$_2$ gas allows a much smoother, faster etch than is possible with Ga$^+$ alone by preventing nanofibre growth under the ion beam. It also protects against contamination with C and O species, passivating the surface with a fluorine-containing species which we deem likely to be GaF$_3$. The combination of spatial resolution, etch rate, low roughness and surface passivation demonstrated by this process should be of great interest to the III-V opto-electronic community.

\section{Acknowledgements}
We thank the staff at \textit{Eblana Photonics Ltd.} and Dr. Megan Canavan, Mr. Clive Downing and Mr. Dermot Daly for technical assistance and fruitful discussions at the Advanced Microscopy Laboratory (AML), CRANN institute, Trinity College Dublin. We acknowledge support from Enterprise Ireland under the Innovation Partnership program [grant: 2017-0542], Science Foundation Ireland [grant: 12/RC/2278] and the Irish Research Council [grant: GOIPG/2014/972].

\FloatBarrier

\bibliography{Mendeley}
\end{document}